\documentclass[aps,prl,twocolumn,groupedaddress,amsfonts,showpacs]{revtex4-1}
\usepackage{graphicx,color}
\usepackage{bm}
\usepackage[latin1]{inputenc}
\usepackage{hyperref}

\begin{document}

\title{Electrical observation of a tunable band gap in bilayer graphene nanoribbons at room temperature}

\author{B. N. Szafranek}
\email{szafranek@amo.de} \homepage{http://www.amo.de}
\author{D. Schall}
\author{M. Otto}
\author{D. Neumaier}
\author{H. Kurz}

\affiliation{Advanced Microelectronic Center Aachen (AMICA), AMO
GmbH, Otto-Blumenthal-Strasse 25, 52074 Aachen, Germany}

\date{\today}

\begin{abstract}
We investigate the transport properties of double-gated bilayer
graphene nanoribbons at room temperature. The devices were
fabricated using conventional CMOS-compatible processes. By
analyzing the dependence of the resistance at the charge neutrality
point as a function of the electric field applied perpendicular to
the graphene surface, we show that a band gap in the density of
states opens, reaching an effective value of $\sim50$~meV. This
demonstrates the potential of bilayer graphene as FET channel
material in a conventional CMOS environment.

\end{abstract}


\maketitle

The two dimensional arrangement of sp$^2$-bonded carbon atoms, known
as graphene, has attracted enormous attention since the group of
Geim pushed it into the spotlight of the solid state community in
2004 \cite{Geim2004}. From a technological point of view the high
mobility and the ampibolar character of charge carriers in graphene
have attracted much interest and have led to investigations on using
graphene as channel material for field-effect transistors (FET)
\cite{Lemme,Lin} and opto-electronic devices \cite{Xia}. However, as
a zero band gap material the intrinsic on/off ratio of graphene
based FETs is rather poor and not suitable for most applications.
Therefore a major focus of research is on the investigation of
opening a band gap in the density of states to increase the on/off
ratio of graphene based FETs.

In bilayer graphene a tunable band gap can be introduced by a
perpendicularly applied electric field. The electric field breaks
inversion symmetry and introduces a band gap at the Dirac point
reaching values of up to 300~meV \cite{McCann,Castro}. Although such
a band gap could be observed by infrared spectroscopy
\cite{Mak,Zhang}, the electrical measurement of such a gap is more
sophisticated. As reported by Oostinga \textit{et al.} an insulating
state occurs in double-gated bilayer graphene, but only at cryogenic
temperatures revealing an effective band gap size below 10 meV
\cite{Oostinga}. They proposed that inter-gap states induced by
inhomogeneities in the local doping lead to a transport mediated via
variable range hopping. Very recently F. Xia \textit{et al.}
\cite{Xia2} managed to detect a band gap in double-gated graphene by
transport measurements at room temperature. By depositing a 9~nm
thick organic layer directly on top of the graphene they could
minimize inhomogeneities in the doping and thus reduce inter-gap
states in the graphene. Our approach for reducing inhomogeneities in
the doping concentration is confining the graphene layer to a
nanoribbon. As recently shown by scanning gate microscopy
micronscale inhomogeneities in the local carrier concentration with
$\Delta n>10^{12}$/cm$^2$ are present in graphene layers
\cite{Malcom}. These inhomogeneities can be avoided by using
nanoscale devices. In such small devices only smaller
inhomogeneities with $\Delta n\sim4\times10^{11}$~/cm$^2$ are
expected on a length scale of typically 20~nm \cite{Deshpande}.

\begin{figure*}
\includegraphics[width=\textwidth]{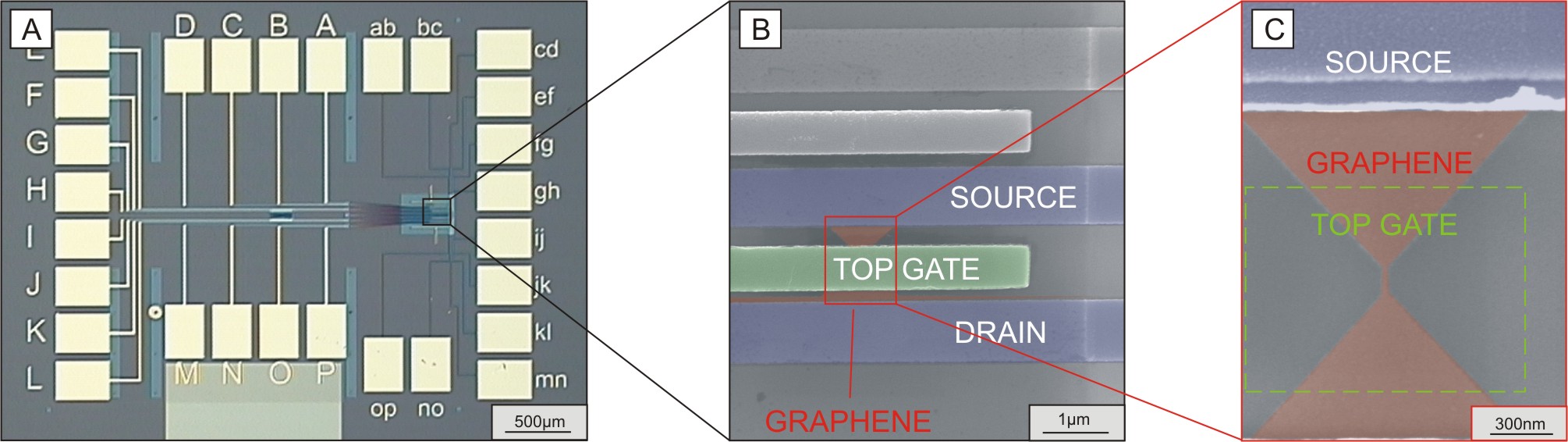}
\caption{Images of double-gated bilayer graphene nanoribbons. A)
Optical image of a sample with a set of 12 individual double-gated
bilayer GNR devices. The 16 source-drain contact pads are labeled from A to
P. The 12 top-gate contact pads are labeled from ab to op. The
SiO$_x$ dielectric is light blue. B) Electron micrograph of one
double-gated bilayer GNR (false color). The ribbon is located below
the top-gate electrode. C) Electron micrograph of a similar bilayer
GNR ($w=30$~nm, $l=250$~nm) before deposition of the top-gate
dielectric and top-gate electrode for illustration of the used
geometry (false color).}
\end{figure*}

For investigating transport in bilayer graphene we fabricated
double-gated graphene nanoribbons (GNRs) using standard
CMOS-compatible processes. Bilayer graphene flakes were deposited by
mechanical exfoliation from natural graphite on highly doped silicon
substrates covered with 90~nm of thermal SiO$_2$. Bilayer flakes
were identified by their optical contrast to the substrate
\cite{Blake}. As back-gate electrode the highly p-doped substrate
was used. In the first lithography step metal contacts were defined
by means of electron beam lithography, thermal evaporation of 20~nm
nickel and a subsequent lift-off. In the next step an etching mask
for the nanoribbons was defined using e-beam lithography and PMMA as
resist. The graphene etching was performed with an oxygen plasma.
With this technique we were able to define nanoribbons with a width
down to 30~nm uncovered by a resist, as PMMA can be removed after
etching. An electron micrograph (false color) of a GNR having a
width of 30~nm and a length of 250~nm is shown in figure 1C. After
patterning the GNR the top-gate dielectric was deposited by thermal
evaporation of $t=25$~nm SiO$_x$ and lift-off. The dielectric
constant of the evaporated SiO$_x$ was measured using standard
MOS-capacitors: $\epsilon_r=6.5\pm0.4$. In the last step the
top-gate electrodes were defined by e-beam lithography and thermal
evaporation of 40~nm nickel followed by a lift-off process. A
finished sample is depicted in figure 1 A-C. The top-gated bilayer
GNRs presented in this paper have a width of 50~nm and a length of
250~nm. All measurements were performed at room temperature (295~K)
in a needle probe station under nitrogen atmosphere. For probing a
HP 4156B semiconductor parameter analyzer was used. Immediately
before the measurement the sample was annealed for one hour in
nitrogen atmosphere at 200 $°$C to minimize contamination.

To investigate the opening of a band gap in bilayer graphene, the
resistance of top-gated bilayer GNRs was measured as a function of
the applied back-gate and top-gate voltage at a fixed source-drain
voltage of 50~mV. In figure 2 the channel resistance of a top-gated
bilayer GNR is depicted as a function of the applied top-gate
voltage for different back-gate voltages ranging from $-40$~V to
40~V. The measured resistance, in
the presented device ranging from $~15$~k$\Omega$ to
$~200$~k$\Omega$, consists of the resistance of the active channel
below the top-gate electrode and the resistance of the leads. As
only the resistance of the active region is required for an
analysis we subtracted the resistance of the relatively long and
narrow leads which is in our case 12~k$\Omega$. This resistance was
estimated using the lead geometry and the corresponding nickel
resistivity and confirmed by a fit of the zero back-gate voltage
transfer-trace similar to the one described in reference \cite{Kim}.
\begin{figure}[b!]
\includegraphics[width=\linewidth]{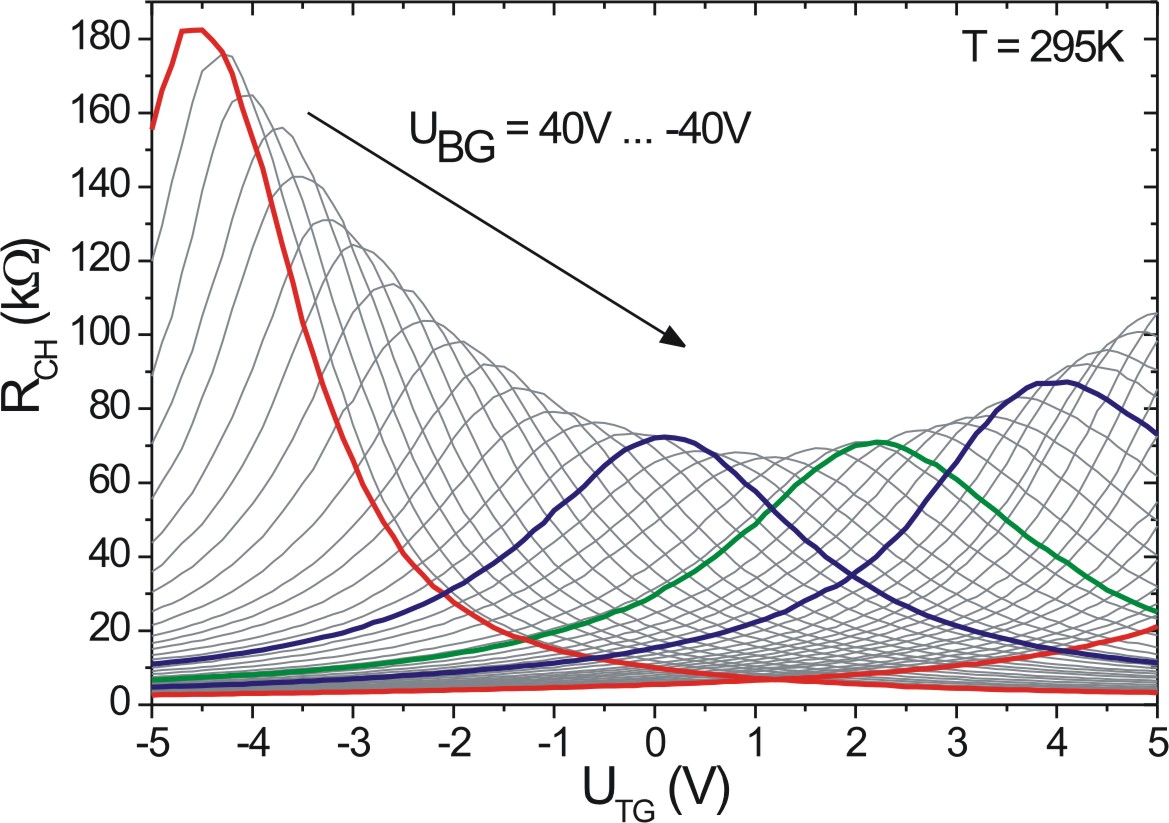}
\caption{Channel resistance of a double-gated bilayer GNR as
function of the top-gate voltage for different back-gate voltages
from -40~V to 40~V (2~V steps). The measurement was performed at
room temperature (295~K) in nitrogen atmosphere at a fixed
source-drain voltage of 50~mV. For clarity the transfer
characteristics at 0~V (green trace), $\pm$10~V (blue traces) and
$\pm$40~V (red traces) are highlighted.}
\end{figure}
With this fit we also extracted a carrier mobility of
$~$1000~cm$^2$/Vs. At zero back-gate voltage (green trace in figure
2) the transfer characteristics show the typical behavior for
bilayer graphene with an on/off ratio slightly below 10. The charge
neutrality point is at a top-gate voltage of 2~V corresponding to a
doping concentration of $2.5\times10^{12}$~/cm$^2$. The resistance
at the charge neutrality point is 71~k$\Omega$, which translates
into a sheet resistance of 6.2~k$\Omega\sim \mathrm{h}/4e^2$. With
increasing back-gate voltage both, the resistance at the charge
neutrality point and the on/off ratio increase, reaching values of
$\sim$180 k$\Omega$ and $\sim$80 at $U_{BG}=40$ V respectively.
These increases reveal the opening of a band gap in the density of
states. The top-gate voltage at the Dirac point is plotted as a
function of the applied back-gate voltage in the inset of figure 3
showing a perfect linear behavior as expected. From the slope the
SiO$_x$ dielectric constant can be estimated to $\epsilon_r=6.4$
which is in good agreement with the $\epsilon_r$ estimated by using
MOS-capacitors.
\begin{figure}
\includegraphics[width=\linewidth]{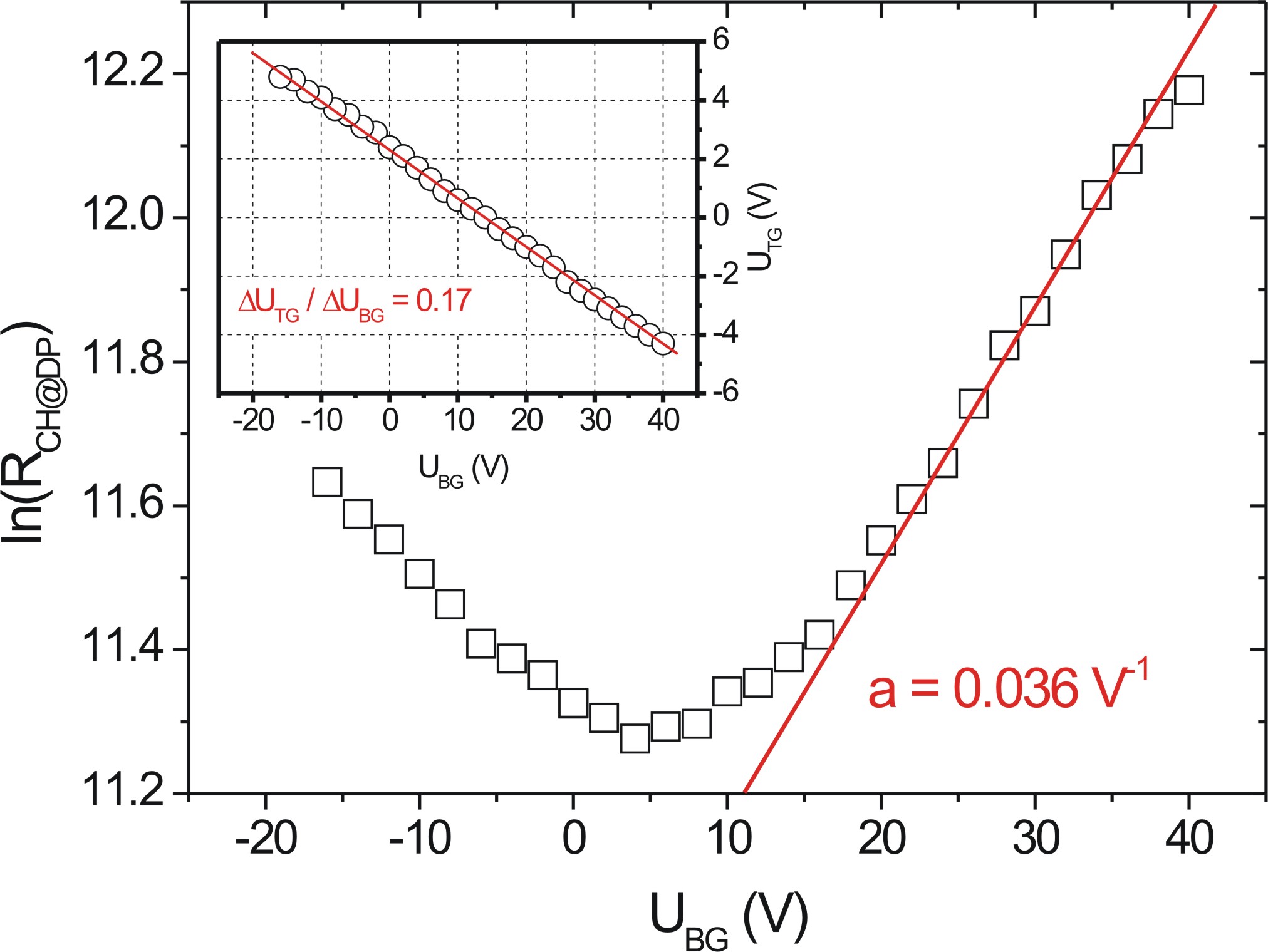}
\caption{Logarithmic channel resistance at the charge neutrality
point of the double gated GNR shown in figure 2 as a function of the
applied back-gate voltage. The straight line gives a slope of
0.036~V$^{-1}$. The inset shows the top-gate voltage of the charge
neutrality point as a function of the applied back-gate voltage. The
straight line is the best linear fit, having a slope of 0.17.}
\end{figure}
To analyze the opening of a band gap more quantitatively we plotted
the logarithm of the channel resistance at the Dirac point
$R_{CH@DP}$ as a function of the applied back-gate voltage in figure
3. In a semiconductor with the Fermi energy located inside the band
gap the resistance $R$ depends on the temperature $T$ and the band
gap energy $E_G$ by $R\propto \mathrm{exp}{(E_G/2k_BT)}$. Assuming
that the band gap vanishes when $R_{CH@DP}$ is minimal, the band gap
energy at $U_{BG}=40$~V can be estimated by:
$E_G(40V)=\mathrm{ln}R_{CH@DP}(40V)-\mathrm{ln}R_{CH@DP}(4V)\sim50$~meV.
As indicated by the straight line in figure 3, the band gap energy
scales roughly linearly for back-gate voltages from 15~V to 40~V, as
one would expect for bilayer graphene at band gap energies smaller
than 150~meV \cite{Zhang, Mak}. The average vacuum displacement
field at the charge neutrality point
$D=(U_{BG}-U_{BG0})\times\epsilon_{r(BG)}/t_{BG}$ is $1.6$~V/nm for
$U_{BG}-U_{BG0}=36$~V; here $U_{BG0}$ is the applied back gate
voltage where $R_{CH@DP}$ is minimal. The estimated band gap energy
is in fair agreement with the value reported by F. Xia \textit{et
al.} (80~meV at $D=1.3$~V/nm \cite{Xia2}), but by a factor of four
smaller than measrued optically \cite{Zhang}. This difference can
most likely be attributed to inter-gap states, which reduce the
electric effective band gap, but do not alter the shape of the
infrared spectroscopy signal. This explanation was already suggested
by J.B. Oostinga \textit{et al.} and confirmed by their analysis of
the temperature dependency of the resistance at the charge
neutrality point, showing an $\mathrm{exp}(1/T^{(1/3)})$ behavior at
cryogenic temperatures, typical for variable range hopping
\cite{Oostinga}. We also investigated electric transport in two
not-annealed top-gated bilayer GNRs having the same geometry as the
device shown above. In these two devices the resistance at the
charge neutrality point also increased with increasing back-gate
voltage, but the effective band gap energy, estimated by the method
shown in this paper, was only $\sim35$~meV. This suggests that
impurities removable by an annealing procedure are a major source of
inter-gap states, which limit the effective band gap opening.

In summary, we have investigated transport in double gated bilayer
graphene nanoribbons at room temperature. The resistance and on/off
ratio of the investigated devices at the charge neutrality point
increases with increasing displacement field showing that a band gap
opens. The maximal achieved effective band gap energy was
$\sim50$~meV at $D=1.6$ V/nm in reasonable agreement with recent
results obtained by Xia \textit{et al.} in a different environment
\cite{Xia2}, but four times smaller than calculated by theory and
measured using infrared spectroscopy \cite{Zhang}. This deviation
between electrical and optical measurements in combination with the
enhancement of the effective gap size by annealing shows that
transport is still limited by randomly distributed inter-gap states
introduced by contamination. Further attempts to increase the
bandgap requires accurate control of the individual process steps to
reduce parasitic contamination.

The authors would like to thank T. Wahlbrink and J. Bolten for
performing the electron-beam lithography. This work was financially
supported by the European Union under contract number 215752
("GRAND") and by the German Federal Ministry of Education and
Research (BMBF) under contract number NKNF 03X5508 ("ALEGRA").



\end{document}